\documentclass[fleqn,twoside,twocolumn,nofootinbib]{revtex4} 

\usepackage{graphicx}
\usepackage{amssymb}
\usepackage{amsmath}

\begin{document}
\title[ strange freeze-out]
{Chemical  Freeze-out of  Strange Particles and Possible Root  of Strangeness Suppression}%

\author{K. A. Bugaev$^{1*}$, D. R. Oliinychenko$^{1}$, J. Cleymans$^{2}$, A.~I.~Ivanytskyi$^{1}$,  
I. N. Mishustin$^{3,4}$
E. G. Nikonov$^{5}$  and V. V. Sagun$^{1}$
}

\mbox{}\\

\affiliation{$^1$Bogolyubov Institute for Theoretical Physics,
Metrologichna str. 14$^B$, Kiev 03680, Ukraine}


\affiliation{$^2$Department of Physics, University of Cape Town, 
Rondebosch 7701, South Africa}

\affiliation{$^3$FIAS,
Goethe-University,  Ruth-Moufang Str. 1, 60438 Frankfurt upon Main, Germany}

\affiliation{$^4$Kurchatov Institute, Russian Research Center, Kurchatov Sqr., Moscow, 123182,  Russia}

\affiliation{$^5$Laboratory for Information Technologies, JINR, Joliot-Curie str. 6, 141980 Dubna, Russia}

\setcounter{page}{1}%
\maketitle


{ \small  
Two approaches to treat the chemical freeze-out of  strange particles in hadron resonance gas model are analyzed. The first one employs their non-equillibration via the usual  $\gamma_s$ factor and such a model describes the hadron multiplicities 
measured in nucleus-nucleus collisions  at AGS, SPS and RHIC energies with $\chi^2/dof \simeq$ 1.15.
Surprisingly, at low energies we find not the strangeness suppression, but its enhancement.
Also  we suggest an alternative approach to treat the strange particle freeze-out separately, but with the full chemical equilibration. This  approach  is based 
on the conservation laws which allow us to 
connect the  freeze-outs of   strange and  non-strange hadrons. 
Within the suggested approach the same set of  hadron  multiplicities can be described better than within  the conventional approach with $\chi^2/dof \simeq $ 1.06.
Remarkably, the fully equilibrated approach describes the strange hyperons and antihyperons much better than the conventional one.}



\vspace*{-5mm}
\section{Introduction} \label{Intro}
\vspace*{-2mm}

Experimental data on multiplicities in heavy ion collisions are traditionally described by the Hadron Resonance Gas Model (HRGM) \cite{Thermal_model_review, KABAndronic:05,KABugaev:Horn2013}. Its core  assumption is that fireball produced in the collision reaches thermal equilibration. 
Using this assumption it is possible 
 to describe the hadronic multiplicities  registered in experiment with the help of two parameters:  temperature $T$ and baryo-chemical potential $\mu_B$. Parameters $T$ and $\mu_B$ obtained from multiplicities fit for different collision energies form the line of chemical freeze-out. In the simplest formulation of the HRGM it is assumed that at this line the inelastic collisions cease simultaneously for all sorts of particles, while to consider the observed  deviation  of  strange particles from the complete chemical equilibrium the additional parameter $\gamma_s$, the strangeness suppression factor, is introduced 
\cite{Rafelsky:gamma}. Although the concept of strangeness suppression proved to be important both in collisions of elementary particles \cite{Becattini:gammaHIC} and in nucleus-nucleus collisions \cite{Becattini:gammaHIC, PBM:gamma} the problem of its justification remains unsolved. Thus, up to now it is unclear what is the main physical reason which is responsible for chemical non-equilibration of strange 
hadrons, since  all  hadrons are in thermal equilibrium at chemical freeze-out and the hadrons built up from $u$ and $d$ quarks do not exhibit  the chemical non-equilibration.  Moreover, as pointed out in \cite{KABAndronic:05} the fit of hadron multiplicities with the strangeness suppression factor $\gamma_s$ improves the quality  of data description, but  still the  fit seldom attains a good quality, especially  at low collision energies. The apparent failure of the $\gamma_s$ fit is clearly seen for the rations of multi-strange baryons $\Xi$ and $\Omega$  at the center of mass energy $\sqrt{s_{NN}} = 8.76,  12.3$ and 17.3 GeV \cite{KABAndronic:05}. Since the multi-strange  baryons are most sensitive to the deviation from chemical equilibrium of strange quarks, but the $\gamma_s$ fit does not 
improve their description sizably, we conclude that there is a different reason for the apparent deviation of strange hadrons from chemical
equilibrium.

In contrast to the $\gamma_s$ concept, here we suggest a modification of HRGM. Instead of a simultaneous  chemical freeze-out  for  all hadrons we consider two different chemical freeze-outs: one for  particles, containing strange charge, even hidden, (we refer to it as strangeness freeze-out, i.e. SFO) and another one (FO) for all other hadrons which contains only $u$ and $d$ (anti)quarks. A partial  justification for such a hypothesis  is given in  \cite{EarlyFO:1,EarlyFO:2,EarlyFO:3}, where the early chemical and kinetic FO of  $\Omega$ hyperons and $J/\psi$ and $\phi$ mesons is discussed for the energies at and  above the highest SPS energy.

One more important feature of the present approach is that FO and SFO parameters are connected by the conservation laws, namely: entropy conservation, baryon charge conservation, strangeness conservation and isospin projection conservation. These laws impose strong restrictions on the fitting parameters. Due to such  restrictions, as we show in the theoretical part, introducing SFO adds only one free parameter for each energy of collision.  Therefore, the number of fitting parameters for the SFO is the same as for the usual HRGM with the strangeness suppression factor $\gamma_s$. Another important feature of the present approach is that we employ the HRGM with multicomponent hard core repulsion \cite{MultiComp:08} which  nowadays provides  the best fit of hadronic multiplicity ratios 
\cite{MultiComp:12} and for the first time it correctly reproduces the energy behavior of  $K^+$ to $\pi^+$ and $\Lambda$ to $\pi^-$ ratios \cite{KABugaev:Horn2013} without spoiling all other hadronic ratios.
 
Note that a similar idea for a separate strangeness FO was recently suggested in  \cite{StrangeFO:13}. We, however, point out that 
our results were presented to the community a few days earlier \cite{NICAWP} and our approach is much more elaborate than the ideal gas treatment of \cite{StrangeFO:13}.

The work consists of two main sections: in section \ref{Model} we give the mathematical formulation of the model and in section \ref{Results} we compare  the fit results for three cases - (i) $\gamma_s$ = 1, no SFO; (ii) $\gamma_s$ is a fitting  parameter, no SFO; (iii) $\gamma_s = 1$, but  SFO is considered. Our conclusions are summarized in section \ref{Conclusions}.

\section{A Brief  Description of Model} \label{Model}

{\bf 1) $\gamma_s$ = 1, no SFO.} We consider multicomponent HRGM, which is currently the best at describing 
the observed hadronic multiplicities. It is the same model as used in \cite{KABugaev:Horn2013}. Hadron interaction is taken into account via hard-core radii, with the different  values for pions, kaons, other mesons and baryons. Best fit values for such radii $R_b$ = 0.2 fm, $R_m$ = 0.4 fm, $R_{\pi}$ = 0.1 fm, $R_K$ = 0.38 fm were obtained in \cite{KABugaev:Horn2013}. The main equations of the model are listed below, but more details of the model can be found in \cite{KABugaev:Horn2013,MultiComp:12}.

Consider  the Boltzmann gas of $N$ hadron species in a volume $V$ that has  the temperature $T$, the baryonic chemical potential $\mu_B$, the  strange chemical potential $\mu_S$ and the chemical potential of the isospin third component $\mu_{I3}$. The system  pressure $p$ and the $K$-th charge density $n^K_i$ ($K\in\{B,S, I3\}$) of the 
i-th hadron sort are given by the expressions  
\begin{eqnarray}\label{EqI}
%
%
\label{EqII}
\frac{p}{T} =  \sum_{i=1}^N \xi_i \,, ~n^K_i = Q_i^K{\xi_i} {\textstyle \left[ 1+\frac{\xi^T {\cal B}\xi}{\sum\limits_{j=1}^N \xi_j} \right]^{-1}} \,, ~ \xi  = \left(
\begin{array}{c}
 \xi_1 \\
 \xi_2 \\
 ... \\
 \xi_s
\end{array}
\right)\,, 
%
\end{eqnarray}
where $\cal B$ denotes a symmetric  matrix of the second  virial coefficients with the elements $b_{ij} = \frac{2\pi}{3}(R_i+R_j)^3$ and 
 the variables $\xi_i$ are the solutions of the following system
\begin{eqnarray}\label{EqIII}
&&\hspace*{-4mm}\xi_i =\phi_i (T)\,   \exp\Biggl[ \frac{\mu_i}{T} - {\textstyle \sum\limits_{j=1}^N} 2\xi_j b_{ij}+{\xi^T{\cal B}\xi} {\textstyle \left[ \sum\limits_{j=1}^N\xi_j\right]^{-1}} \Biggr] \,, \quad \quad \\
&&\hspace*{-4mm}\phi_i (T)  = \frac{g_i}{(2\pi)^3}\int \exp\left(-\frac{\sqrt{k^2+m_i^2}}{T} \right)d^3k  \,.
\label{EqIV}
\end{eqnarray}
Here   the full chemical potential of the $i$-th hadron sort $\mu_i \equiv Q_i^B \mu_B + Q_i^S \mu_S + Q_i^{I3} \mu_{I3}$ is expressed in terms of the corresponding charges $Q_i^K$  and their  chemical potentials,  $ \phi_i (T) $ denotes 
the thermal particle  density of  the $i$-th hadron sort of mass $m_i$ and degeneracy $g_i$, and  $\xi^T$  denotes  the row of  variables $\xi_i$.  
Therefore, the main fitting 
parameters are temperature $T$, baryonic chemical potential $\mu_B$ and the chemical potential of the third projection of  isospin  $\mu_{I3}$, whereas the strange chemical potential  $\mu_S$ is found from the vanishing strangeness condition.

Width correction is taken into account by averaging all expressions containing mass by Breit-Wigner distribution having a  threshold. The effect of resonance decay $Y \to X$ on the final hadronic multiplicity is taken into account as $n^{fin}(X) = \sum_Y BR(Y \to X) n^{th}(Y)$, where $BR(X \to X)$ = 1 for the sake of convenience. 
The masses, the  widths and the strong decay branchings of all hadrons  were  taken from the particle tables  used  by  the  thermodynamic code THERMUS \cite{THERMUS}.

{\bf 2) $\gamma_s$ is a fitting  parameter.} In this case we follow the conventional way of introducing $\gamma_s$ and replace $\phi_i$ in Eq. (\ref{EqII}) as
\begin{eqnarray} \label{eq:gamma_s}
\phi_i(T) \to \phi_i(T) \gamma_s^{s_i} \,,
\end{eqnarray}
where $s_i$ is number of strange valence quarks plus number of strange  valence anti-quarks.

{\bf 3) SFO.} Let us consider two freeze-outs instead of one. The strangeness chemical  freeze-out  is assumed to occur for all strange particles  at the temperature $T_{SFO}$, the baryonic chemical potential $\mu_{B_{SFO}}$,  
the isospin
third projection chemical potential $\mu_{I3_{SFO}}$ and the   three dimensional space-time extent (effective volume) of the freeze-out hypersurface  $V_{SFO}$. The  freeze-out of hadrons which are built of the $u$ and $d$ quarks, i.e. FO, is assumed to be described by  its own  parameters $T_{FO}$, $\mu_{B_{FO}}$, $\mu_{I3_{FO}}$, $V_{FO}$. Eqs. (\ref{EqI})--(\ref{EqIV}) for FO and SFO remain the same as for a simultaneous  FO of all particles.
In both cases $\mu_S$ is found from the net zero strangeness condition. 
The major difference of the SFO 
 is in the conservation laws and the corresponding modification  of the resonance decays. 
Thus, we assume that between two freeze-outs  the system is sufficiently dilute
and hence its evolution is governed by the continuous hydrodynamic evolution which conserves 
the entropy. Then 
equations for  the entropy, the baryon charge and  the isospin projection conservation connecting two freeze-outs are as follows:
\begin{eqnarray}
s_{FO} V_{FO} = s_{SFO} V_{SFO} \,, \label{ent_cons}\\
n^B_{FO} V_{FO} = n^B_{SFO} V_{SFO} \,, \label{B_cons}\\
n^{I_3}_{FO} V_{FO} = n^{I_3}_{SFO} V_{SFO} \,. \label{I3_cons}
\end{eqnarray}
Getting rid of the effective  volumes we obtain
\begin{eqnarray} 
\label{Eq:FO_SFO1}
\frac{s}{n^B} \biggl|_{FO} = \frac{s}{n^B} \biggr|_{SFO} \,,  \quad  \frac{n^B}{n^{I_3}} \biggl|_{FO} = \frac{n^B}{n^{I_3}} \biggr|_{SFO}
%
 \,.  
\end{eqnarray}
Therefore, the variables $\mu_{B_{SFO}}$ and $\mu_{I3_{SFO}}$ are not free parameters, since
 they are found from the system  (\ref{Eq:FO_SFO1})  and only $T_{SFO}$ should be fitted.
Thus, for SFO the number of independent fitting parameters is the same as in the case of $\gamma_s$ fit. 

The number of resonances  appeared due to decays are considered as follows:
\begin{eqnarray} \label{Eq:SFO_decays}
%
\frac{N^{fin}(X)}{V_{FO}} = \sum_{Y \in FO} BR(Y \to X) n^{th}(Y) + \nonumber \\
 \sum_{Y \in SFO} BR(Y \to X) n^{th}(Y) \frac{V_{SFO}}{V_{FO}} \,. 
\end{eqnarray}
Technically this is done by multiplying all the thermal concentrations for SFO by $n^B_{FO}/n^B_{SFO} = V_{SFO}/V_{FO}$ and applying the conventional resonance decays. 
\section{Results} \label{Results}
{\bf Data sets and fit procedure.} In our choice of the data sets we basically  followed Ref. \cite{KABAndronic:05}. Thus, at  the AGS energy range of collisions ($\sqrt{s_{NN}} = 2.7 -4.9$ GeV) the data are  available for the kinetic beam energies from 2 to 10.7 AGeV.  For the beam energies 2, 4, 6 and 8 AGeV there are only a few data points available: the yields for pions \cite{AGS_pi1, AGS_pi2}, for protons \cite{AGS_p1, AGS_p2}, for kaons  \cite{AGS_pi2} (except for 2 AGeV), for  $\Lambda$ hyperons the integrated over $4 \pi$ data are available \cite{AGS_L}. For the beam  energy 6 AGeV there exist  the $\Xi^-$ hyperon  data integrated over $4 \pi$ geometry   \cite{AGS_Kas}. However, the data for  the  $\Lambda$ and $\Xi^-$ hyperons
have to be corrected \cite{KABAndronic:05}, and instead of the raw experimental data we used their  corrected values of  Ref. \cite{KABAndronic:05}.
For the highest AGS center of mass energy $\sqrt{s_{NN}} = 4.9$ GeV (or the beam energy 10.7 AGeV) in addition
to the mentioned data for pions, (anti)protons and  kaons  there exist data for $\phi$ meson \cite{AGS_phi},
for  $\Lambda$ hyperon \cite{AGS_L2} and  $\bar \Lambda$ hyperon \cite{AGS_L3}.
Similarly to \cite{KABugaev:Horn2013}, here  we analyzed  only  the  NA49  mid-rapidity data   \cite{KABNA49:17a,KABNA49:17b,KABNA49:17Ha,KABNA49:17Hb,KABNA49:17Hc,KABNA49:17phi}.
Since  the RHIC high energy  data of different collaborations agree with each other, we  analyzed  the STAR results  for  
$\sqrt{s_{NN}}= 9.2$ GeV \cite{KABstar:9.2}, $\sqrt{s_{NN}}= 62.4$ GeV \cite{KABstar:62a},
 $\sqrt{s_{NN}}= 130$ GeV \cite{KABstar:130a,KABstar:130b,KABstar:130c,KABstar:200a} and  200 GeV \cite{KABstar:200a,KABstar:200b,KABstar:200c}. 
To simplify the numerical efforts and to avoid considering the effective  volumes we fit particle ratios rather than the multiplicities. The best fit criterion is a minimality of $\chi^2 = \sum_i  \frac{(r^{theor}_i - r^{exp}_i)^2}{\sigma^2_i} $, where $r_i^{exp}$ is an experimental value of i-th particle ratio, $r_i^{theor}$ is our prediction and $\sigma_i$ is a total error of experimental value.

{\bf Fit with $\gamma_s$.} 
Inclusion of $\gamma_s$ is expected to improve the description of ratios containing the strange particles. It may also give room in parameter space that will ultimately lead to improvement of ratios that contain no strange particles. In our investigation we pay a special attention to the $K^+/\pi^+$ ratio, because it is  usually considered as the most problematic one for HRGM.

\begin{figure}[Htbp]
    \begin{center}
    \includegraphics[height=70mm]{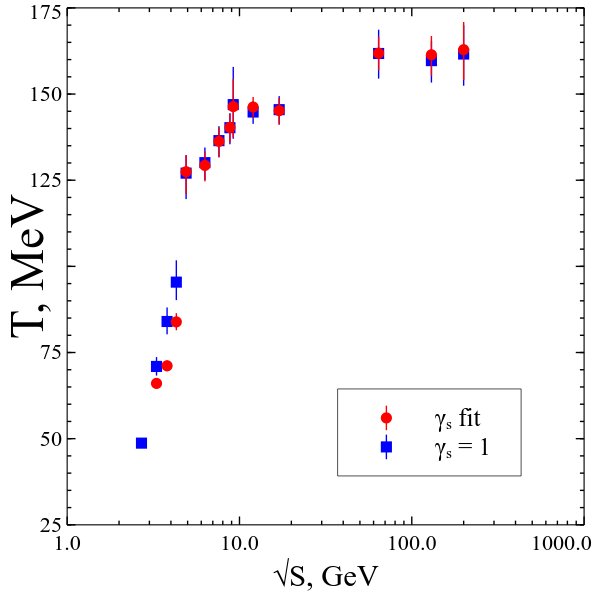}
    \end{center}
   \begin{center} 
   \includegraphics[height=70mm]{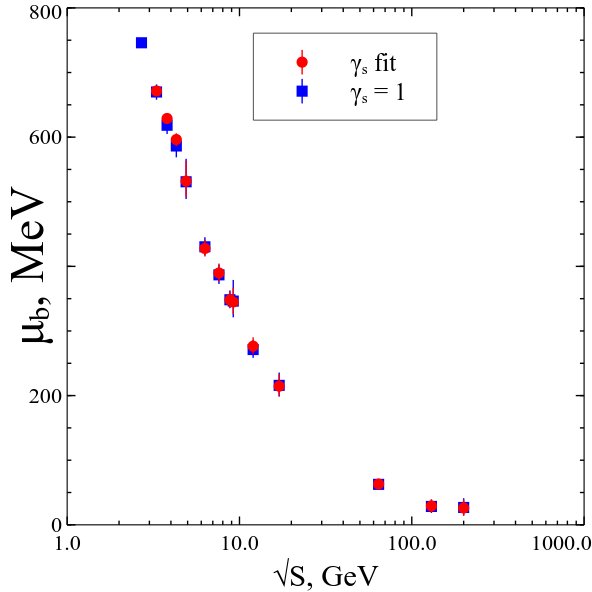}
   \end{center}   
   \begin{center} 
   \includegraphics[height=70mm]{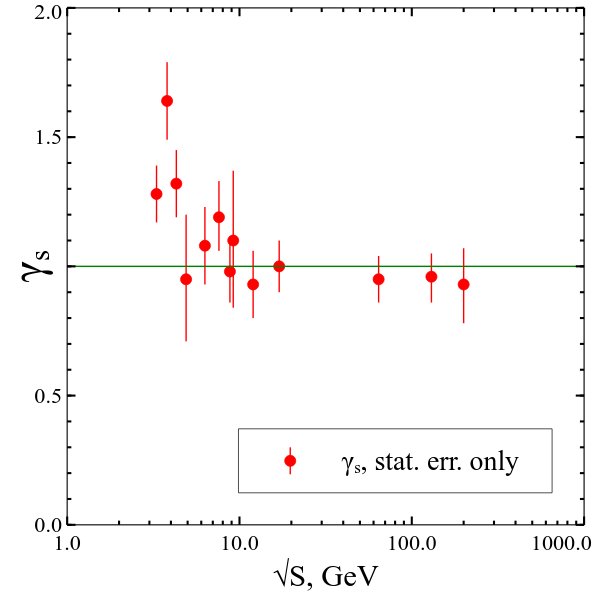}
   \end{center}   
\vspace*{-8.4mm}
 \caption{Behavior of parameters for the  $\gamma_s$  fit and for a single chemical FO with $\gamma_s$ = 1. Upper panel: temperature $T$. Middle panel: baryo-chemical potential $\mu_B$. Lower panel: $\gamma_s$.}
  \label{Fig:gs_param}
\end{figure}

The behavior of fit parameters  is shown in  Fig. \ref{Fig:gs_param} for  $T$, $\mu_B$ and $\gamma_s$. The obtained values of the chemical FO temperatures and baryo-chemical potentials in the case with $\gamma_s$ fit do not considerably differ from the case, when $\gamma_s$ = 1, while   the behavior of $\gamma_s(\sqrt{s_{NN}})$ 
demonstrates entirely new results.
Thus, at low energies for $\gamma_s$ fit we found not a suppression, but a strangeness enhancement, i.e. $ \gamma_s > 1$.
These findings  are in a drastic contrast to the results of the statistical hadronization model \cite{Becattini:gammaHIC} for the 
fit of hadronic multiplicities, measured in nuclear collisions. The $\gamma_s$ values  reported in  \cite{Becattini:gammaHIC}
demonstrate a suppression, i.e. $ \gamma_s < 1$, for all ASG and SPS energies.  We, however, note that the fit quality 
of hadronic multiplicities reported  in  \cite{Becattini:gammaHIC}  is essentially worse, compared even to the present model without $\gamma_s$ fit, and, therefore, one cannot rely on the  statistical hadronization model conclusions on the  $\gamma_s$ 
values. 

For 14 values of collision energy
$\sqrt{s_{NN}} = $  2.7, 3.3, 3.8, 4.3, 4.9, 6.3, 7.6, 8.8, 9.2, 12, 17, 62.4, 130, 200 GeV
the best description with $\gamma_s$ fit gives  $\chi^2/dof$ = 63.4/55 = 1.15, which is only a very slight improvement compared to the  results $\chi^2/dof$ = 80.5/69 = 1.16 found for a single chemical freeze-out with $\gamma_s=1$. Note, however, that the value of $\chi^2$ itself, not divided by number of degrees of freedom,  has improved notably. This fact motivates us to study what ratios are  improved.

\begin{figure}[Htbp]
    \begin{center}
    \includegraphics[height=70mm]{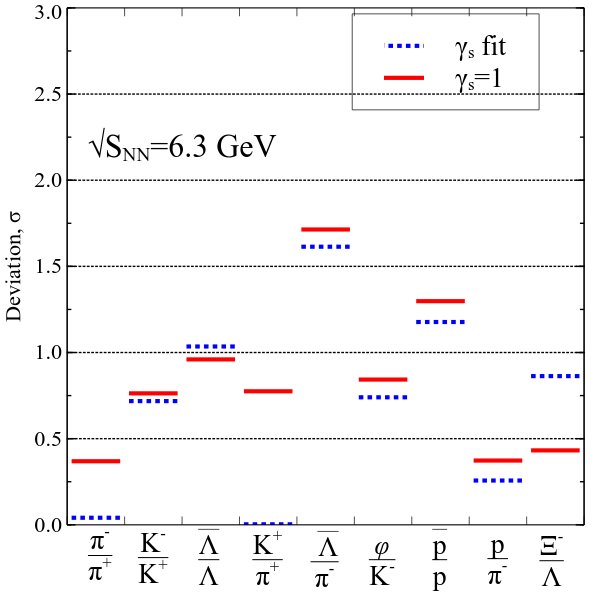}
    \end{center}
   \begin{center} 
   \includegraphics[height=70mm]{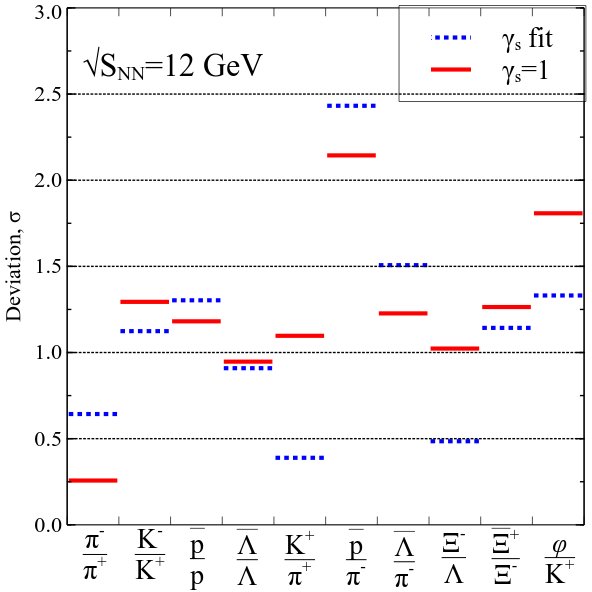}
   \end{center}   
   \begin{center} 
   \includegraphics[height=70mm]{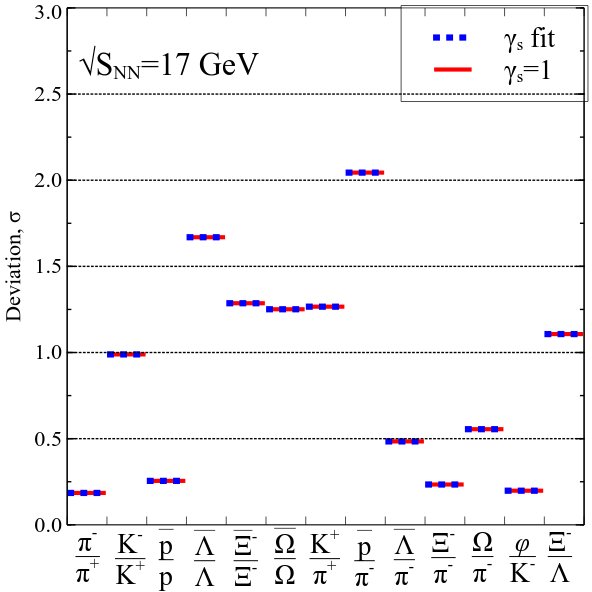}
   \end{center}   
\vspace*{-8.4mm}
 \caption{Relative deviation of theoretical description of ratios from experimental value in units of experimental error $\sigma$. The symbols on OX axis demonstrate the particle ratios. OY axis shows $\frac{|r^{theor} - r^{exp}|}{\sigma^{exp}}$, i.e. the modulus of  relative deviation for $\sqrt{s_{NN}}$ = 6.3, 12 and 17 GeV. 
  Solid lines correspond to the model with a single  FO of all hadrons and $\gamma_s =1$, while the dashed lines correspond to the model with $\gamma_s$ fit.} 
  \label{Fig:dev_G}
\end{figure}

At AGS energies $\sqrt{s_{NN}}$ = 2.7, 3.3, 3.8 and 4.3 GeV the  number of available ratios  is small (4, 5, 5, 5 respectively) and  only kaons and $\Lambda$ contain strange quarks. Since  the data  description  is rather good even within the ideal gas model \cite{KABAndronic:05}, the inclusion of  $\gamma_s$ into a fit does   improve the fit quality, but it
 leads to the vast minima in the parameter space and large errors of $\gamma_s$. Moreover,  at low energies the fit is unstable: two local minima with very close $\chi^2$ are often found. For instance,   for  $\sqrt{s_{NN}}=$  3.8 GeV  we find 
 $\gamma_s \simeq 1.6$ in the deepest minimum,  while   in another  minimum, next to the deepest one,   $\gamma_s \simeq 0.8$. 
An existence of two local minima with  close values of $\chi^2$  at $\sqrt{s_{NN}} = 2.7-4.3$ GeV  
 tells us that the $\gamma_s$ concept has to be improved further in order to resolve this problem.

\begin{figure}[t]
\includegraphics[width=63mm]{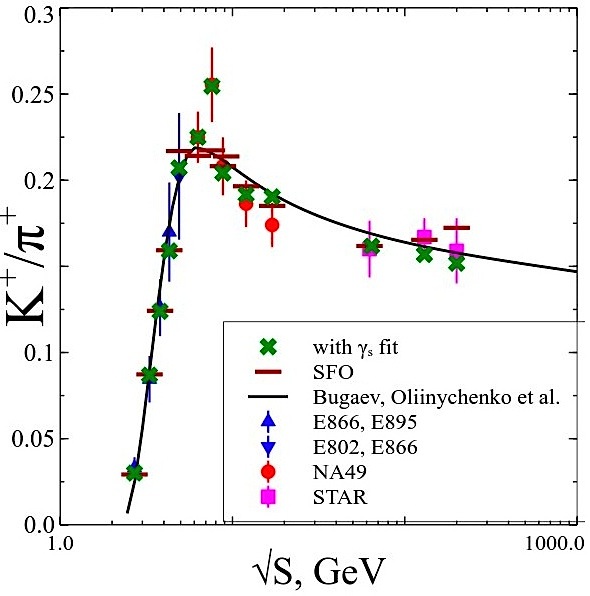}
\vspace*{-4.4mm}
\caption{Description of ${K^+}/{\pi^+}$ ratio. Solid line is the result of \cite{KABugaev:Horn2013}. Crosses stand for the case with $\gamma_s$ fitted, while the horizontal bars correspond to SFO.}
\label{Fig:gs_horn}
\end{figure}

 At $\sqrt{s_{NN}} =$ 4.9 GeV $\gamma_s$ does not improve ratios description, but its behavior  is stable and hence  $\gamma_s = 1$  within the error bars. At $\sqrt{s_{NN}} =$ 6.3-12 GeV $K^+/\pi^+$ ratio is notably improved, while the description of other ones has improved only slightly or even became worse (see the typical examples  in the upper and middle  panels of Fig. \ref{Fig:dev_G}). At  $\sqrt{s_{NN}} \ge $ 17  GeV energies there is no special improvement. Conclusively, fitting $\gamma_s$ provides an opportunity to improve the Strangeness Horn description to $\chi^2/dof = 3.3/14$, i.e.   better than it was done in \cite{KABugaev:Horn2013} with $\chi^2/dof = 7.5/14$. 
  The Strangeness Horn itself is shown in  Fig. \ref{Fig:gs_horn}. We would like to stress  that even the highest point of the Horn is reproduced now, that makes our theoretical horn as sharp as an experimental one. However, the overall $\chi^2/dof  \simeq 1.15$ obtained for  $\gamma_s$ fit is only slightly better compared to the result $\chi^2/dof  \simeq 1.16$ found in   \cite{KABugaev:Horn2013}. Moreover,  $\gamma_s$ fit does not essentially improve the ratios with strange baryons and, hence,   we consider an alternative approach.

\begin{figure}[t]
\includegraphics[width=63mm]{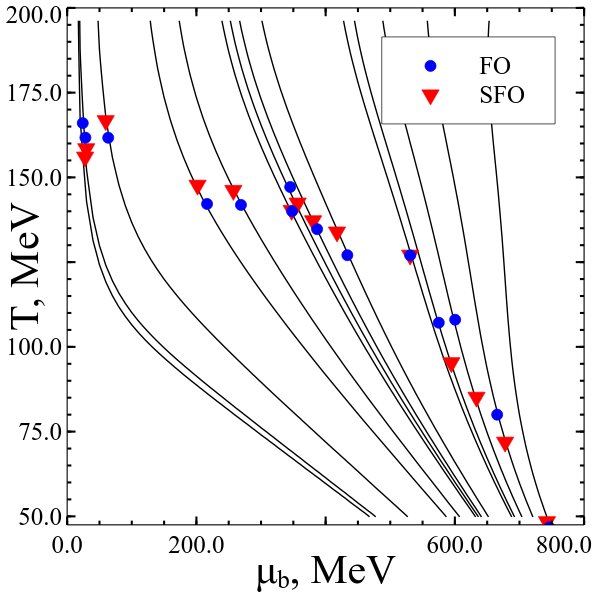}
\vspace*{-4.4mm}
\caption{Points of chemical freeze-outs in the model with two freeze-outs.  Triangles correspond to SFO, their coordinates are ($\mu_{B_{SFO}},\,T_{SFO}$), while circles  correspond to FO and their coordinates are ($\mu_{B_{FO}},\,T_{FO}$). The curves  correspond to isotherms  ${s}/{\rho_B} = const$ connecting two freeze-outs.
}
\label{Fig:FO_SFO_param}
\end{figure}

\begin{figure}[Htbp]
    \begin{center}
    \includegraphics[height=70mm]{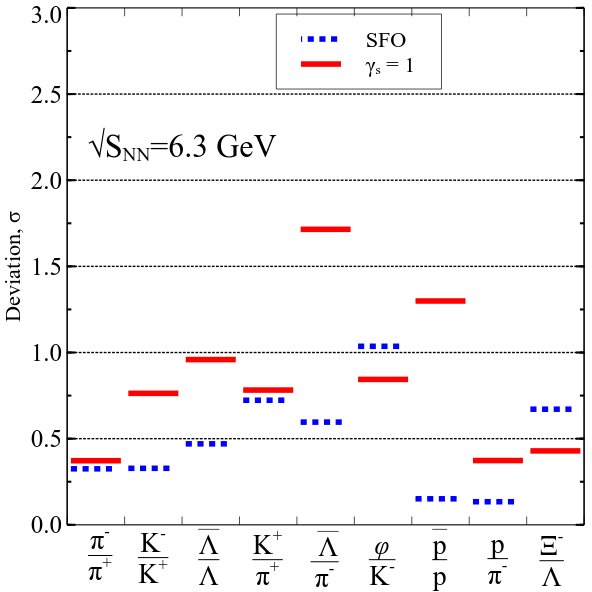}
    \end{center}
   \begin{center} 
   \includegraphics[height=70mm]{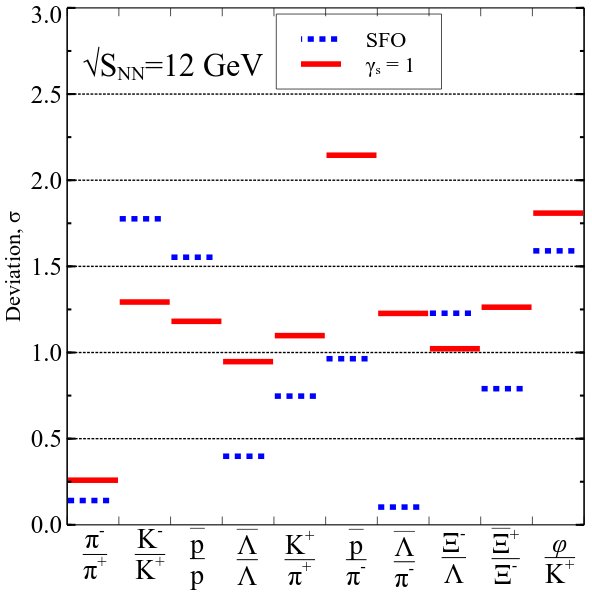}
   \end{center}   
   \begin{center} 
   \includegraphics[height=70mm]{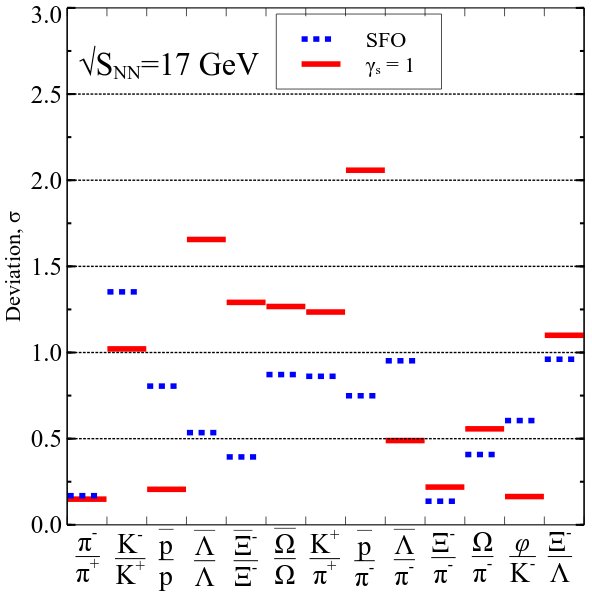}
   \end{center}   
\vspace*{-6.4mm}
 \caption{Same as in Fig. \ref{Fig:dev_G}. Solid lines correspond to model without SFO and $\gamma_s = 1$, dashed lines correspond to model with SFO.} 
  \label{Fig:dev_FO_SFO}
\end{figure}

{\bf Fit with SFO and no $\gamma_s$.} In this case  $\gamma_s$ = 1 is fixed for all energies, but  FO and SFO parameters are connected by  conservation laws (\ref{Eq:FO_SFO1}). Therefore for SFO 
at each collision energy there is only one fitting parameter, namely
$T_{SFO}$, while 
 other parameters are found from the system  (\ref{Eq:FO_SFO1}). 
Like  in the previous case we study two things: behavior of parameters and what  ratios are  improved.
First of all we found out that for SFO case $\chi^2/dof = 58.5/55 \simeq 1.06$, which means that an improvement due to SFO introduction (global $\chi^2/dof$: 1.16 $\to$ 1.06) is sizably  better, than an improvement due to $\gamma_s$ fitting (global $\chi^2/dof$: 1.16 $\to$ 1.15).
For low energies the  situation is similar to the previous case. At  $\sqrt{s_{NN}} = $ 2.7, 3.3, 3.8, 4.3 and 4.9 GeV the original  description obtained within the multicomponent model \cite{KABugaev:Horn2013} is very good and hence it is not improved significantly, but, at least, it is not worse than  the description   obtained by $\gamma_s$ fit. 
Similar results are found at highest RHIC energies $\sqrt{s_{NN}} >   62.4$ GeV. 
As one can see from Fig. \ref{Fig:FO_SFO_param} for these two energy domains  the SFO temperature is below the  FO temperature. 
At intermediate  energies we see a systematic improvement of ratios description. Three plots corresponding to collision energies at which an improvement after SFO introduction is the most significant, $\sqrt{s_{NN}}$ = 6.3, 12 and 17 GeV,  are shown in Fig. \ref{Fig:dev_FO_SFO}. 
As one can see from Fig. \ref{Fig:dev_FO_SFO} for $\sqrt{s_{NN}}$ = 6.3, 12 and 17 GeV  the SFO approach greatly  improves all the ratios with more than one $\sigma$ deviation.  For  $\sqrt{s_{NN}}$ = 6.3 GeV the SFO greatly   improves 
$\bar \Lambda/\pi^-$ and $\bar p/p$ ratios. For $\sqrt{s_{NN}}$ = 12 GeV  four ratios out of  eight   with more than one $\sigma$ deviation, namely  $K^+/\pi^+$, $\bar \Lambda/\Lambda$, $\bar \Lambda/\pi^-$ and  $\bar \Xi^+/\Xi^-$ are sizably   improved. The data 
measured at $\sqrt{s_{NN}}$ = 17 GeV were not improved by  $\gamma_s$ fit at all, while the SFO approach allows us 
to greatly  improve the fit quality.  Fig. \ref{Fig:dev_FO_SFO} clearly demonstrates that due to SFO fit 
the seven out of eight problematic ratios of $\gamma_s =1$ fit  moved from the region of  deviation exceeding   $\sigma$ to the region of deviations being smaller than $\sigma$. The most remarkable of them are $\bar p/\pi^-$, $\bar \Lambda/\Lambda$  $\bar \Xi^-/\Xi^-$ and  $\bar \Omega/\Omega$. Thus, a separation of FO and SFO relaxes the strong connection  between the non-strange and strange baryons  and allows us for the first time not only to correctly describe   the ratios of strange antibaryons to the same strange baryons, but also it allows us to successfully reproduce the antiproton to pion ratio.
As it is seen from Fig. \ref{Fig:gs_horn} the SFO fit quality  is  worse compared to the Strangeness Horn  fit by $\gamma_s$, but overall  it is very good with $\chi^2/dof = 6.3/14$. 
\section{Conclusions} \label{Conclusions}
In this paper we performed a high quality fit of the hadronic multiplicity ratios measured at AGS, SPS and RHIC energies. In contrast to earlier  beliefs established on the low  quality fit \cite{Becattini:gammaHIC}, we find that within the error bars  in heavy ion collisions  there is a sizable enhancement of strangeness, i.e. $\gamma_s > 1$, 
at $\sqrt{s_{NN}} = $  2.7, 3.3, 3.8,  4.9, 6.3, 9.2 GeV. 
However, the effect of apparent strangeness enhancement can be  successfully explained by the idea of separate chemical freeze-out of all strange hadrons. Our analysis shows that for the same number of  fitting parameters the SFO approach is working not worse  than the $\gamma_s$ approach, but for $\sqrt{s_{NN}} = $ 6.3, 12 and 17 GeV  it improves the fit quality tremendously. At these energies we see that  $\bar p/\pi^-$ and  strange antibaryons to same strange baryon ratios are much better described than within the $\gamma_s$ approach. 
This allows us to conclude that    an  apparent strangeness enhancement  is  due to 
the separate strangeness chemical freeze-out. 
\mbox{}\\
\hfill\\
\noindent
{\bf Acknowledgments.}
The authors appreciate the valuable comments of  M. Bleicher, A. S. Sorin and  G. M. Zinovjev. 
K.A.B., D.R.O., A.I.I. and V.V.S.   acknowledge  a partial  support of the Program `On Perspective Fundamental Research in High Energy and Nuclear Physics' launched by the Section of Nuclear Physics  of NAS of Ukraine. 
K.A.B.  and I.N.M.  acknowledge a partial support provided by the Helmholtz 
International Center for FAIR within the framework of the LOEWE 
program launched by the State of Hesse. 
The work of  E.G.N.  was supported in part by the Russian
Foundation for Basic Research, Grant No. 11-02-01538-a.


\end{document}